\documentclass[prb,aps,twocolumn,draft,tightenlines,%
showpacs,floatfix,superscriptaddress]{revtex4} \usepackage{psfig}
\usepackage{amssymb}

\begin{document}
\title{Spin oscillations in transient diffusion of a spin pulse in $n$-type semiconductor quantum wells} 
\author{M. Q. Weng}
\affiliation{Structure Research Laboratory, University of Science \&%
  Technology of China, Academia Sinica, Hefei, Anhui, 230026, China}
\affiliation{Department of Physics, University of Science \&%
  Technology of China, Hefei, Anhui, 230026, China}%
\altaffiliation{Mailing Address.}  \author{M. W. Wu}
\thanks{Author to whom correspondence should be addressed}%
\email{mwwu@ustc.edu.cn}%
\affiliation{Structure Research Laboratory, University of Science \&%
  Technology of China, Academia Sinica, Hefei, Anhui, 230026, China}
\affiliation{Department of Physics, University of Science \&%
  Technology of China, Hefei, Anhui, 230026, China}%
\altaffiliation{Mailing Address.}  \author{Q. W. Shi}
\affiliation{Structure Research Laboratory, University of Science \&%
  Technology of China, Academia Sinica, Hefei, Anhui, 230026, China}

\date{\today}
\begin{abstract}
  By studying the time and spatial evolution of a spin-polarized pulse 
in $n$-type semiconductor quantum wells, we highlight
  the importance of the off-diagonal spin coherence in spin diffusion
  and transport.  Spin oscillations and spin polarization reverse
  along the the direction of spin diffusion in the absence of the
  applied magnetic field are predicted from our investigation.
\end{abstract}
 \maketitle

There are growing numbers of experimental and theoretical
investigations on spin related phenomena due to potential applications
in semiconductor spintronic devices such as spin filters and spin
transistors.\cite{wolf,spintronics,folk_2003,zhou_2004,%
datta_1990,johnson_1993,monsma_1998} 
An important prerequisite for the realization of such devices is to
understand how 
the spin-polarized electrons transport from one position to another.
In experiment, coherent spin propagation over a long distance has been
reported.\cite{kikkawa2} Spin injections from ferromagnetic materials
into semiconductors have been extensively
investigated.\cite{hanbicki_2003} In theory most works are based on
the quasi-independent electron model  and are focused on the
diffusive transport regime\cite{spintronics,schmidt} in which the spin
polarization of the current is controlled by the longitudinal spin
dephasing through a relaxation time approximation.
There are some efforts on investigating the spin dephasing in the spin
transport: Takahashi {\it et al.} calculated the spin diffusion
coefficients by solving the kinetic equations with only the
electron-electron scattering.\cite{taka_prb_1999} Bournel {\it et
al.}\cite{bournel_ssc_1997} and Saikin {\it et al.}\cite{saikin_2003}
studied the spin transport in semiconductor heterostructures
with the Rashba effect and the electron-phonon scattering by Monte
Carlo simulation.  However, in these investigations either the
off-diagonal inter-spin-band correlations $\rho_{{\bf
k}\sigma\,-\sigma}=\langle c^{\dag}_{{\bf k}\sigma} c_{{\bf
k}\,-\sigma}\rangle$ are discarded (or not explicitly included) or the
spin dephasing is simply introduced through the relaxation time
approximation.

Recently we performed a many-body investigation of spin transport in
GaAs (100) quantum wells (QW's) by self-consistently solving the
many-body kinetic transport equations together with the Poission
equation.  We have explicitly taken into account the spatial
inhomogeneity, the D'yakonov-Perel' (DP) mechanism and all
scattering.\cite{weng_prb_2002} In our theory, both the diagonal
intra-spin-band correlations, {\em i.e.}, the electron distribution
function $f_{{\bf k}\sigma}=\langle c^{\dag}_{{\bf k}\sigma}c_{{\bf
k}\sigma}\rangle$, and the off-diagonal inter-spin-band correlations
$\rho_{{\bf k}\sigma\,-\sigma}$ are explicitly included. The spin
dephasing time, the spin/charge diffusion length together with the
mobility are calculated self-consistently.  We further pointed out a
novel spin dephasing mechanism in the spin diffusion/transport that
the inhomogeneous broadening due to the interference between the
electrons/spins with different momenta along the direction of the
diffusion can cause spin dephasing in the presence of the scattering
and the resulting dephasing can be more important than the dephasing
due to the DP mechanism.  In this paper, we further investigate the
time evolution of a spin-polarized pulse (SPP) in $n$-type GaAs
(100) QW through the many-body kinetic Bloch equations.  We highlight
the importance of the {\em off-diagonal spin correlations} in the spin
diffusion and transport.

In $n$-type GaAs QW's, the dominant spin dephasing mechanism is the DP
mechanism.\cite{dp,meier} By taking account of the DP term, the
kinetic Bloch equations can be written as
\begin{widetext}
  \begin{equation}
    \label{eq1}
    {\partial\rho({\bf R},{\bf k},t)\over\partial t} -{1\over 2}
    \bigl\{{\bf \nabla}_{\bf R}\bar{\varepsilon}({\bf R},{\bf k}, t),
    {\bf \nabla}_{\bf k}\rho({\bf R},{\bf k},t)\bigr\} +{1\over 2}
    \bigl\{{\bf \nabla}_{\bf k}\bar{\varepsilon}({\bf R},{\bf k},t),
    {\bf \nabla}_{\bf R}\rho({\bf R},{\bf k},t)\bigr\}
    -\left.{\partial\rho({\bf R},{\bf k},t)\over\partial
    t}\right|_{c}\nonumber\\ =\left.{\partial\rho({\bf R},{\bf
    k},t)\over\partial t}\right|_{s}.
  \end{equation}
\end{widetext}
\noindent Here $\rho({\bf R},{\bf k}, t)$ represents a single particle
density matrix. The diagonal elements describe the electron
distribution functions $\rho_{\sigma\sigma}({\bf R},{\bf
k},t)=f_{\sigma}({\bf R},{\bf k},t)$ of wave vector ${\bf k}$ and spin
$\sigma$($=\pm 1/2$) at position ${\bf R}$ and time $t$. The
off-diagonal elements $\rho_{\sigma-\sigma}({\bf R},{\bf k}, t)$
describe the inter-spin-band correlations (coherences) for the spin
coherence. The quasi-particle energy
$\bar{\varepsilon}_{\sigma\sigma^{\prime}}({\bf R},{\bf k},t)$, in the
presence of a moderate magnetic field ${\bf B}$ and with the DP
term\cite{dp} included, reads $
\bar{\varepsilon}_{\sigma\sigma^{\prime}}({\bf R},{\bf k},t)=
\varepsilon_{k}\delta_{\sigma\sigma^{\prime}} +\bigl[g\mu_B{\bf
B}+{\bf h}({\bf k})\bigr]\cdot
\mbox{\boldmath$\sigma$\unboldmath}_{\sigma\sigma^{\prime}}/ 2 -
e\psi({\bf R},t) +\Sigma_{\sigma\sigma^{\prime}}({\bf R},{\bf k},t)$,
where $\varepsilon_k=k^2/2m^\ast$ is the energy spectrum with $m^\ast$
denoting the electron effective
mass. $\mbox{\boldmath$\sigma$\unboldmath}$ are the Pauli
matrices. ${\bf h}({\bf k})$ denotes the effective magnetic field from
the DP effect which contains contributions from both the
Dresselhaus\cite{dress} term and the Rashba term.\cite{ras} In this
paper, we only consider the Dresselhaus term which can be written
as\cite{eppenga} $h_x({\bf k})=\gamma k_x(k_y^2-\kappa_z^2)$ and
$h_y({\bf k})=\gamma k_y(\kappa_z^2-k_x^2)$ with $\gamma$ denoting the
spin-orbit coupling strength\cite{aronov} and $\kappa_z^2$
representing the average of the operator $-(\partial/\partial z)^2$
over the electronic state of the lowest subband.  The electric
potential $\psi({\bf R},t)$ satisfies the Poission equation
\begin{equation}
  \label{poission}
  {\bf \nabla}_{\bf R}^2\psi({\bf R},t)= -{e\bigl[n({\bf
  R},t)-n_0({\bf R})\bigr]/\epsilon}\ ,
\end{equation}
where $n({\bf R},t)=\sum_{\sigma{\bf k}}f_{\sigma}({\bf R},{\bf k},t)$
is the electron density at position ${\bf R}$ and time $t$ and
$n_0({\bf R})$ is the positive background electric charge density.
$\Sigma_{\sigma\sigma^{\prime}} ({\bf R}, {\bf k}, t) =-\sum_{{\bf
q}}V_{{\bf q}} \rho_{\sigma\sigma^{\prime}}({\bf R},{\bf k}-{\bf
q},t)$ is the Hartree-Fock self-energy, with $V_{\bf q}$ standing for
the Coulomb matrix element.  In 2D case, $V_{\bf q}$ is given by
$V_{\bf q} = 2\pi e^2/[\epsilon_0(q+\kappa)]$, with
$\kappa=(2e^2m^{\ast}/\epsilon_0)\sum_{\sigma}f_{\sigma}({\bf k}=0)$
being the inverse screening length. $\epsilon_0$ represents the static
dielectric constant.  It is noted that if one only takes account of
the diagonal elements $\rho_{\sigma\sigma}$ and neglects all the
off-diagonal elements $\rho_{\sigma-\sigma}$ in Eq.\ (\ref{eq1}), the
first three terms on the left hand side of the equation correspond to
the driving terms in the classical Boltzmann equation, modified with
the DP term and the selfenergy term from the Coulomb Hartree
contribution.  ${\partial \rho({\bf R},{\bf k},t) \over \partial
t}|_c$ and ${\partial \rho({\bf R},{\bf k},t) \over \partial t}|_s$ in
the Bloch equations (\ref{eq1}) are the coherent and scattering terms
respectively. The coherent term describes the electron spin precession
around the applied magnetic field and the effective magnetic field
from the DP term. Its expression as well as that of the scattering
term $\left.{\partial\rho({\bf R},{\bf k},t)\over\partial
t}\right|_{s}$ are given in detail in Refs. \onlinecite{weng_prb_2002}
and \onlinecite{c0302330}.

It is noted that by using a spin-flip time $\tau_{\text{sf}}$ to
describe the spin dephasing caused by the DP term and summing over the
momentum, one is able to derive the diffusion equations for charge and
spin densities of electrons in the so called ``mean field''
approximation.\cite{spintronics,qi_prb_2003} However, as pointed out
in our previous papers,\cite{weng_prb_2002} the adoption of the ``mean
field'' approximation removes the interference between the electrons
with different momentums and thus overlooks the inhomogeneous
broadening that causes additional spin dephasing in the spin
transport.  We further point out in this paper that by using the
spin-flip time approximation, some of the most marked features of the
DP mechanism are thrown away.

We assume that at initial time $t=0$ there is a SPP centered at
$x=0$. The electrons are locally in equilibrium, {\em i.e.},
$f_{\sigma}(x,{\bf k},0)=\{\exp\bigl[(\epsilon_k-\mu_{\sigma}(x))/T_e
\bigr]+1\}^{-1}$, where $\mu_{\sigma}(x)$ stands for the chemical
potential of electrons with spin $\sigma$ at position $x$ and is
determined by the corresponding electron density: $N_{\sigma}(x, 0) =
\sum_{k} f_{\sigma}(x,{\bf k},0)$. The shape of the initial spin pulse
is assumed to be Gaussian like:
\begin{equation}
  \Delta N(x,0) = N_{\frac{1}{2}}(x,0) - N_{-\frac{1}{2}}(x,0)=\Delta
  N_0 e^{-x^2/\delta x^2},
  \label{deltan}
\end{equation}
with $\Delta N_0$ and $\delta x$ representing the peak and width of
the SPP respectively. We further assume that there is no spin
coherence at the initial time, $\rho_{\sigma-\sigma}(x,{\bf k}, 0)=0$.
This SPP can be achieved by a circularly polarized laser pulse.

\begin{figure}[htb]
  \centering
  \psfig{figure=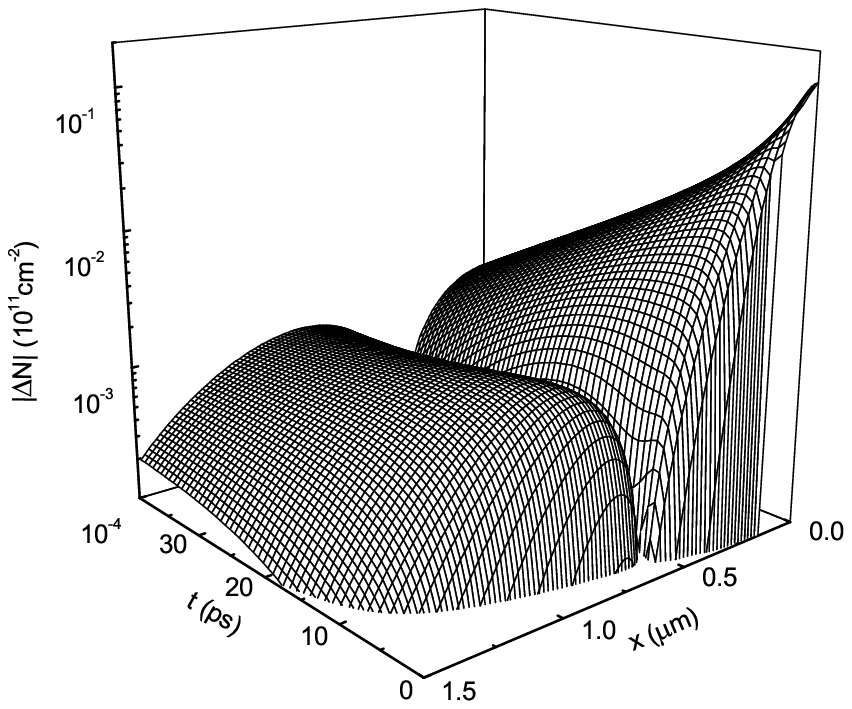,width=8.1cm,height=7.cm,angle=0} \vskip
  -1cm \psfig{figure=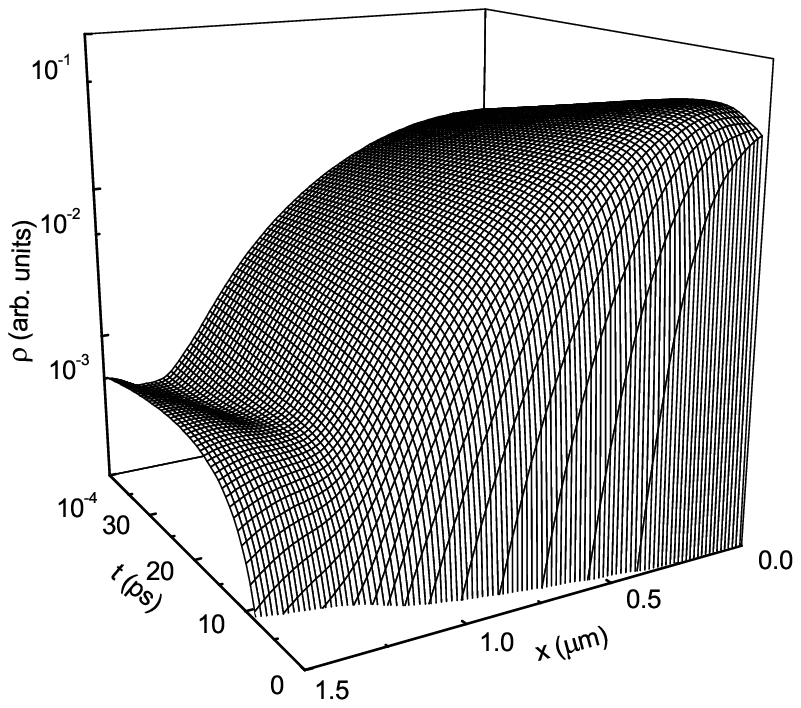,width=8.1cm,height=7.cm,angle=0}
  \caption{The absolute value of the spin imbalance 
    $|\Delta N|$ and the incoherently summed spin coherence $\rho$
    vs. the position $x$ and the time $t$ for the impurity free
    ($N_i=0$) case.}
  \label{fig1}
\end{figure}

By numerically solving the kinetic equations (\ref{eq1}) together with
the initial conditions and the Poission equation (\ref{poission}), we
are able to study the temporal evolution of the SPP with $T=200$~K and
$B\equiv 0$. A typical calculation of the spin diffusion in $n$-type
GaAs QW is carried out by choosing the total electron density
$N_e(\equiv n_0({\bf R}))= 4\times 10^{11}$~cm$^{-2}$, the maximum
spin imbalance in Eq.\ (\ref{deltan}) $\Delta N_0=1\times
10^{11}$~cm$^{-2}$ and the width of the spin pulse $\delta
x=0.15~\mu$m.  The material parameters are taken from
Ref.~\onlinecite{made}. The absolute value of the spin imbalance
$|\Delta N(x,t)|$ and the incoherently summed spin
coherence\cite{wu_epjb_2000} $\rho = \sum_{\bf k}|\rho_{{\bf
k}\sigma\,-\sigma}({\bf R}, t)|$ are plotted as functions of the
position $x$ along the diffusion direction and the time $t$ in
Fig.~\ref{fig1} (a) and (b) respectively for impurity free case
($N_i=0$).  Here $N_i$ denotes the impurity density.  One can see from
the figure that due to the strong diffusion as well as the spin
dephasing, the spin polarization at the center of the spin pulse
decays very fast initially.  In the mean time, the spin coherence
$\rho$ goes up due to the precession of spins in the presence of the
effective magnetic field of the DP term.  After 10~ps, as the
diffusion becomes weaker due to the smaller spatial gradient, the
decay rate at the center of the spin signal slows down and the spin
coherence begins to decay due to the diffusion as well as the spin
dephasing.  It is noted that the spin polarization away from the
center, {\em e.g.} in the region $0.12$~$\mu$m~$<x<0.15~\mu$m, first
increases due to the net spin diffusion from the center and then
decays after the diffusion from the center becomes moderate. For the
region out of the initial spin pulse ($x>0.15$\ $\mu$m), the combined
effect of the diffusion and the dephasing leads to more complicated
behaviors.  The most striking feature of the evolution of the SPP is
that the spin polarization can be {\em opposite} to the initial one
even in the {\em absence} of the applied magnetic field and there are
oscillations in the time evolution of the spin polarization at some
positions.

\begin{figure}[htb]
  \centering \psfig{figure=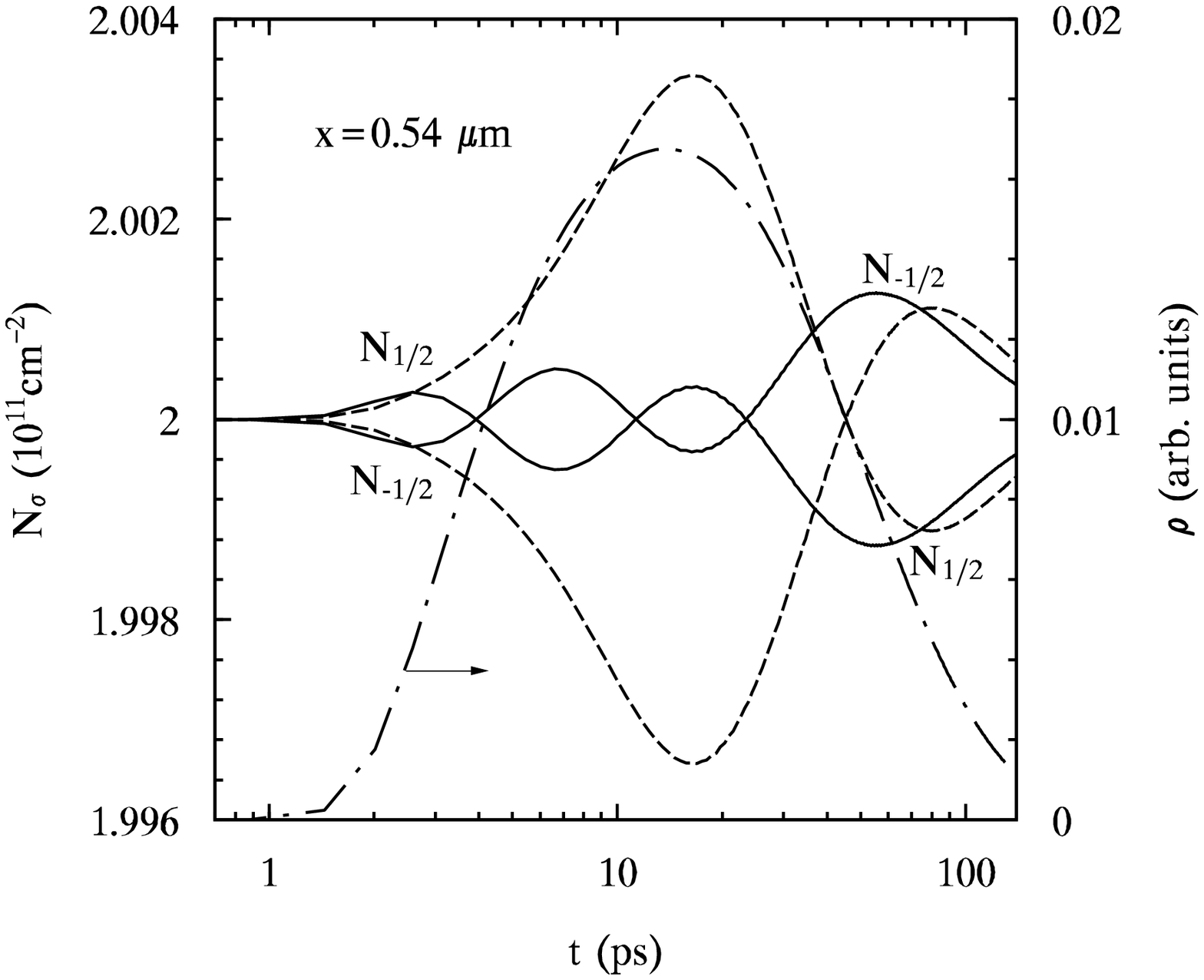,width=8cm,height=7cm,angle=0}
  \psfig{figure=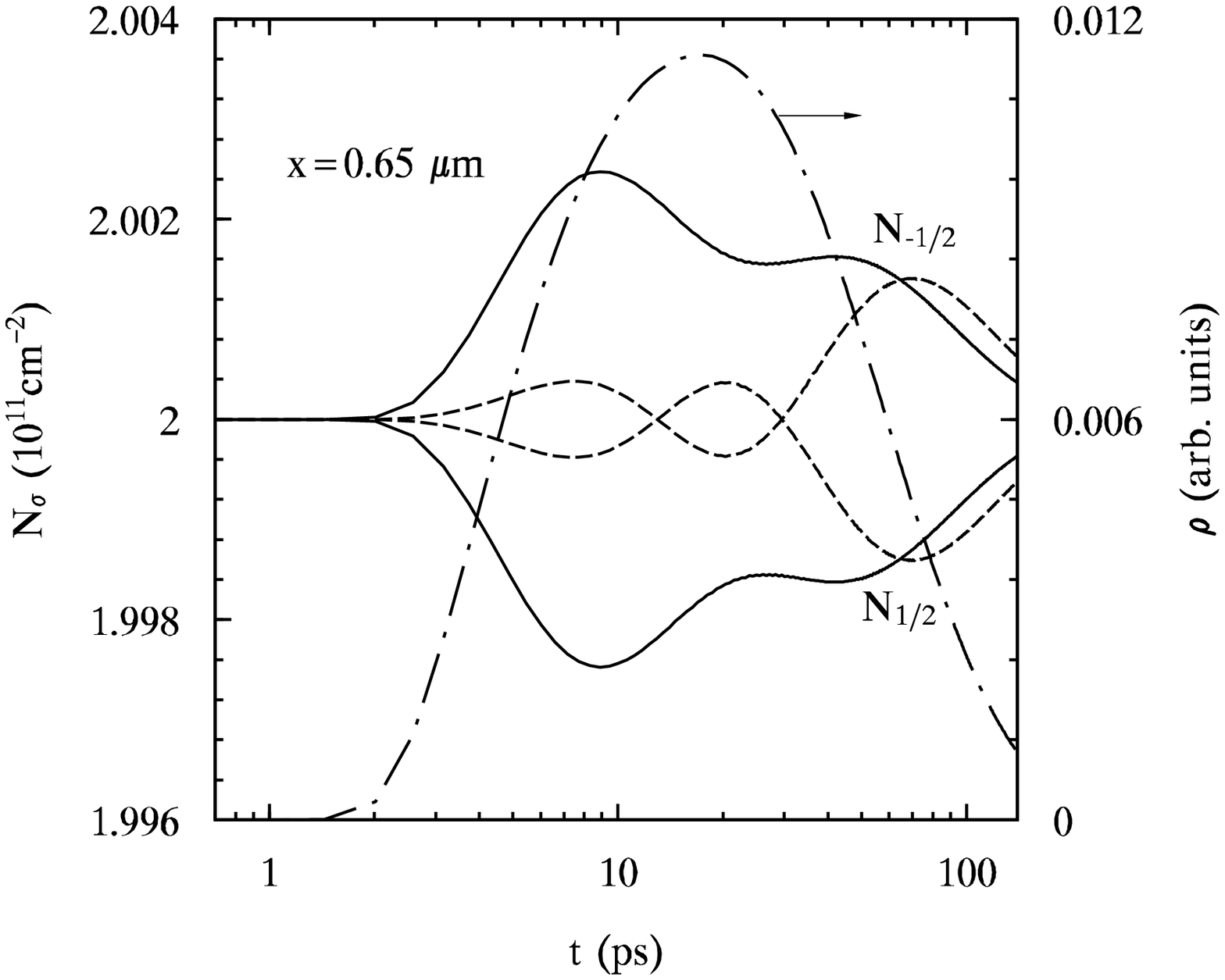,width=8cm,height=7cm,angle=0}
  \caption{The electron densities for different spins at (a) $x=0.54~\mu$m
    and (b) $x=0.65~\mu$m vs. time $t$ with solid curves for $N_i=0$
    and dashed ones for $N_i=0.1~N_e$. The dotted curves are the
    corresponding incoherently summed spin coherence at the same
    position for $N_i=0$ case.  }
  \label{fig2}
\end{figure}

From Fig. 1(a) one can see that there is another peak in the spin
polarization at the positions out of the initial spin pulse after
10~ps. However the spin polarization of this second peak is {\em
opposite} to the initial one. As the time goes on, the second peak
becomes larger due to the decay of the first peak in the center. After
30~ps, the heights of these two peaks are comparable. Moreover, in the
region $0.5~\mu$m$< x<0.7~\mu$m, there are oscillations in the time
evolution of spin polarization. The details are shown in
Fig.~\ref{fig2} where the densities of the electrons with different
spin $N_\sigma(x,t)$ are plotted at two typical positions $x=0.54$ and
0.65\ $\mu$m for the cases with (dashed curves) and without (solid
curves) impurities.

It is seen from Fig. 2(a) that at $x=0.54$\ $\mu$m for impurity free
case, there is no spin signal in the first picosecond as the position
is located out of the initial spin pulse. Then the spin signal starts
to build up due to the diffusion of both the diagonal electron
distribution $f_\sigma(x,{\bf k},t)$ and the off-diagonal spin
coherence $\rho_{{\bf k}\sigma -\sigma}(x,t)$ from the center.  With
the joint effects of the DP term as well as the diffusion, the spin
signal reaches its first peak at about 2.5~ps and then decreases. It
gets a crossing of spin-up and -down electron densities at 4~ps. After
that the spin-down electrons exceed the spin-up ones and hence the
spin polarization changes sign.  The difference between the spin-up
and -down electrons further increases until 7~ps when the difference
arrives at another peak.  The oscillations go on with the periods
become larger and larger. After the third oscillation, the period
becomes too long to observe any oscillation in the time regime of our
calculation. As a result after the last crossing at 25~ps, the spin
density of spin-down electrons is larger than that of the spin-up ones
and the spin polarization is reversed.  One can further see from the
figure that, in the regime when the polarization oscillates, the spin
coherence keeps increasing. This indicates that the spin coherence
$\rho$ at this position mainly comes from the diffusion from the
center.  It is seen from the coherent terms of the Bloch equations
that the imbalance of the distribution functions with the opposite
spin and the spin coherence can transfer to each other by the DP
effective magnetic field.  Therefore, the additional spin coherence
$\rho$ diffused from the center induces the spin polarization
oscillations at position outside the initial spin pulse.

It is further noted that the oscillations differ in positions. For
example, as shown in Fig.~\ref{fig2}(b), at $x=0.65$~$\mu$m (and
positions farther away), the oscillations are almost marginal but the
feature that the spin polarization after a long time is opposite to
the initial one remains.  Moreover it is noted that in Fig.\ 2(a),
when the spin pulse arrives, its polarization points to the initial
one. Nevertheless, in Fig.\ 2(b) for positions farther away from the
initial pulse, it is interesting to see that the spin signal arrives
with spin polarization opposite to the initial one. This is understood
that at positions far away from the initial pulse, what diffused from
the center of the spin pulse is the off-diagonal spin coherence and
the diagonal spin imbalance is quite small already at the edge of the
initial pulse and therefore its diffusion to the outer space is
marginal. The off-diagonal spin coherence induces the reverse of the
spin polarization as what said above.  This can be seen from the fact
that for $t<10$\ ps in Fig.\ 2(b), both diagonal and off-diagonal
terms increase with time.  For positions farther away from the center,
the arrival of the spin polarization also includes the diagonal
componants but with the opposite spin polarization.

We now explore the effect of the impurity to the spin diffusion.  In
Fig.\ 2 we also plot the corresponding curves of electron densities
with impurity density $N_i=0.1~N_e$ as dashed curves.  The figure
shows that the impurities dramatically change the behavior of the spin
diffusion but do not eliminate the spin oscillations and the reverse
of the spin polarization when time is long enough.  By comparing the
solid curves with the dashed ones it is noted that when time is long
enough, say around 100\ ps, the spin polarization for the case with
impurities is always larger than that without impurities. This is
understood that for the impurity free case, the mobility of electrons
is larger and spin polarization is easier to diffuse away from the
center.  Moreover, the non-magnetic impurities tend to retain the spin
coherence. This is because the impurity scattering drives the
electrons to a homogeneous state in the momentum space and therefore
counters the effect of the DP term that drives the electrons to an
inhomogeneous state and leads to spin
dephasing.\cite{meier,wu_epjb_2000,c0302330}

A similar spin oscillation without an applied magnetic field has been
reported by Brand {\em et al.}.\cite{brand_2002} Nevertheless it is
noted that it is quite different from what discussed here. In the work
of Brand {\em et al.}, the spin oscillation happens in a spacial
homogeneous system. And the oscillation is due to the breakdown of the
assumption of the collision domination in the spin dephasing at
extreme low temperature (a few Kelvin). For temperature higher than
10\ K, the oscillation disappears as the electron collision rate
increases and the assumption of the collision domination recovers.  In
our work, the spin oscillation comes from the diffusion (spacial
gradient) and happens {\em outside} the initial spin pulse at very
high temperature (200\ K). The origin of the spin oscillation is
therefore by no means only due to the small electron collision rates
as in the case of Brand {\em et al.}, but due to the combination of
the diffusion and the precession of spin signals: At positions just
outside of the initial spin pulse, the off-diagonal spin coherences of
electrons with large momentums first reach there. As the DP term is
proportional to the momentum, this fraction of electrons also share
large precession frequencies.  Therefore it is possible for the spin
signal to oscillate at first few picoseconds and in the region just
outside spin pulse when the collisions do not affect the momentum of
electrons dramatically. As time passes by, the electrons with
relatively smaller momentums, consequently with smaller precession
frequencies, also arrive at the region where the oscillation occurs.
Therefore due to the joint effects of the diffusion as well as the
increasing impact of the collisions, the oscillation becomes slower
and slower and totally vanishes after a few hundred picoseconds. For
the spin signals in the center of the initial pulse there is no spin
oscillation for the temperature of our investigation.  This coincides
with the results of Brand {\em et. al.}\cite{brand_2002} as well as
what we discovered in high temperature cases.\cite{c0302330}

The reverse and oscillation of the spin polarization of a spin pulse
along the diffusion in the absence of the applied magnetic field can
only be achieved when {\em both} the diagonal and off-diagonal terms
of electron density matrix and the precession caused by the DP term
are considered. Once the relaxation time approximation is adopted to
describe the effect of the DP term or the off-diagonal term is
dropped, the spin signal at the positions outside the initial SPP
first increases due to the diffusion from the spin pulse then
decreases monotonically due to the diffusion as well as the dephasing
and the spin polarization never changes the sign. This is because in
the framework of the relaxation time approximation, the most important
difference between the spin dephasing caused by the DP mechanism and
that caused by the other mechanisms is wiped off. Other spin dephasing
mechanisms such as the Elliot-Yafet mechanism\cite{elliot} and the
Bir-Aronov-Pikus mechanism \cite{bap} cause the spin dephasing through
instantaneous spin-flip scattering. However, the DP term acts as a
momentum dependent magnetic field around which the electron spins
precess. This precession results in an inhomogeneous broadening in
spin phases and leads to spin dephasing in the presence of the
scattering. In additional to the dephasing, it is also possible that
the precession may switch the magnetic momentum in the transport even
without the applied magnetic field and in the high temperature regime.
For some special positions, the combined effect of the diffusion and
the precession may lead the spin to oscillate as shown in
Fig.~\ref{fig2}.

In conclusion, we perform a study of transient spin diffusion of a SPP
in $n$-type GaAs QW's by self-consistently solving the full kinetic
Bloch equations in which both the diagonal terms, the distribution
functions, and the off-diagonal terms, the inter-spin-band spin
coherences, of the density matrix are included. We show that by taking
the effect of the DP term to the diagonal and off-diagonal terms of
the density matrix into account, the spin polarization outside the
initial SPP can be reversed even without magnetic field.  Moreover,
for some special positions, the spin signals oscillate with time. The
reverse and oscillations of the spin signals in spin
diffusion/transport at positions outside the initial spin pulse have
not been reported either theoretically or experimentally for $n$-type
semiconductors and are understood from the diffusion of the
off-diagonal terms of the electron density matrix which oscillates due
to the precession of electron spins around the effective magnetic
field due to the DP effect.  We also stress that these features can
only be achieved when the {\em off-diagonal terms} of the electron
density matrix are included {\em explicitly} in the theory.

We would like to thank S. Zhang for his critical reading of this 
manuscript. MWW is supported by the ``100 Person Project'' of 
Chinese Academy of Sciences and Natural Science Foundation 
of China under Grant No. 90303012. MQW is partially supported 
by China Postdoctoral Science Foundation.


\end{document}